**Title:** Micro- and Nano-scale Measurement Methods for Phase Change Heat Transfer on Planar and Structured Surfaces

**Authors:** Jacopo Buongiorno, David G. Cahill, Carlos H. Hidrovo, Saeed Moghaddam, Aaron J. Schmidt, Li Shi

## Abstract

In this opinion piece, we discuss recent advances in experimental methods for characterizing phase change heat transfer. We begin with a survey of techniques for high resolution measurements of temperature and heat flux at the solid surface and in the working fluid. Next, we focus on diagnostic tools for boiling heat transfer and describe techniques for visualizing the temperature and velocity fields as well as measurements at the single bubble level. Finally, we discuss techniques to probe the kinetics of vapor formation within a few molecular layers of the interface. We conclude with our outlook for future progress in experimental methods for phase change heat transfer.

## 1. Introduction

There has been increasing interest in the potential for nanostructured surfaces to enhance phase change heat transfer, as discussed in several other papers in this issue. Heat transfer between a solid and a working fluid is fundamentally an interface phenomenon, and advancing both scientific understanding and engineering practice in this field requires experimental capabilities for high resolution characterization of the thermodynamics and kinetics of the underlying interfacial processes.  The science and engineering discipline of phase-change heat transfer is not alone in facing this daunting challenge.  Tribology, corrosion, water purification, electrochemical energy conversion, catalysis, photocatalysis, crystal growth, and adhesion of contacting surfaces are just some of the current problems in science and engineering that share these challenges. The traditional science of solid surfaces focuses on clean crystals in ultra-high vacuum, where powerful tools based on electron scattering and photoemission can be used to precisely probe structure and composition of a surface with high spatial and temporal resolutions.  In the topical and more practical problems listed above, co-existence of multiple phases makes the ambient highly complex, the surfaces of interest are often inhomogeneous on multiple length scales, and elevated pressures and temperatures of the vapors and liquids involved greatly restrict the applicability of most experimental tools for the measurement of phase, composition, temperature, and heat flux.

Despite these challenges, there has been substantial work on experimental techniques to study different aspects of phase change heat transfer across many length and time scales, and the resulting data has shed light on many of the complex physical processes involved. In this opinion piece, we discuss recent advances in experimental methods for characterizing several aspects of phase change heat transfer from the nanoscale to the macroscale. We begin with a discussion on measurement challenges in phase change heat transfer in the context of the problem of nucleate boiling, and survey some techniques for high resolution measurements of temperature and heat flux at the solid surface and in the working fluid. We consider recent experiments based on microfabricated heater and sensor arrays, as well as a variety of optical techniques for visualizing the solid surface and the temperature and velocity fields in the adjacent fluid. Next we discuss recent microscale measurements at the single bubble level during boiling, and then go on to describe the use of photothermal and photoacoustic techniques to probe the kinetics of vapor formation within a few molecular layers of the interface. We conclude with our outlook for future progress and directions.

## 2. Measurement needs and existing approaches

The phenomenon of nucleate boiling highlights the experimental challenges associated with measuring phase change processes over multiple time and length scales. As a critical mechanism of heat transfer in numerous practical applications ranging from large-scale power plants to small-scale heat sinks, nucleate boiling has been studied for several decades. However, its physical nature is still plagued with large uncertainties, due in large part to the complex coupling of mass, momentum, and energy transport that occurs between the solid surface, the wetting liquid, and the vapor produced to generate the bubble. The complexity of the problem and the lack of physical understanding of the process have resulted in diverse hypotheses concerning the physics of the involved heat transfer processes [73-76]. For example, the microlayer [73,74] presence and its level of contribution to the bubble growth and the overall surface heat transfer is still a subject of debate and some studies (e.g. [75]) suggest that most heat transfer to the liquid takes place via a transient conduction process triggered by the bubble growth and departure.

Evaluation of analytical models requires new, highly-resolved data for the many individual (sub)phenomena that constitute what we call "boiling". In particular, a complete characterization of nucleate boiling would require knowledge of bubble size and shape throughout the ebullition cycle, bubble departure frequency (including growth and wait times), nucleation site density, wetted area fraction (i.e. fraction of the boiling surface that is in contact with the liquid phase), temperature field (2D on the boiling surface, and D within the fluid), velocity field, local heat flux, and evolution of the microlayer at the base of the bubble. To aid in the development and validation of models and simulations, the above quantities should be measured over a meaningful range of conditions; in dimensionless terms: $P/P_{cr}$ (pressure), Bo (heat flux), Ja (subcooling) and Re (flow rate in flow boiling).

Traditionally, boiling heat transfer has been studied with fairly simple apparatuses, consisting of a heater (typically a metallic plate or rod) instrumented with thermocouples, and high-speed cameras for visualization of the interactions among the fluid phases and the boiling surface. Thermocouples can only measure temperature at discrete locations on the boiling surface, thus little information on the temperature distribution can be obtained, while the usefulness of high-speed video is typically limited by poor optical access to the nucleation site and interference from adjacent bubbles, although the total reflection approach [35-38] allows for visualization of the boiling process from below the boiling surface through a transparent heater at any conditions.

Because no one diagnostic can generate all the desired data, a requirement key to the success of any modern study of boiling is combination/synchronization of multiple diagnostics. Second-generation two-phase flow diagnostics, such as multi-sensor conductivity and optical probes [39,40] and wire-mesh probes [41], can measure bubble diameter and velocity near the boiling surface. However, these approaches are intrusive, and also produce data only at discrete locations within the boiling fluid. On the other hand, X-ray and $\gamma$-ray tomography is non-intrusive, but rather costly/cumbersome as the radiation source has to be rotated at high speed around the test section, which also may limit the time and/or space resolution of the technique; ultra-fast approaches have been developed to increase the time and space resolution in recent years, including 2D non-tomographic approaches that can reach extraordinarily high spatial resolution (<1 $\mu$m), but require massive X-rays sources [42]. Several modern and powerful experimental approaches are coming to full maturity for use in two-phase systems, and can be used to satisfy the data needs discussed above. For example, infrared-based visualization of thermal patterns on the boiling surface was pioneered by Theofanous et al. [43] in the early 2000s, and then adopted by many others [38, 44-48], while arrays of micro-heaters, individually controlled to achieve a constant temperature boundary condition, can be used to resolve the local instantaneous surface heat flux, as shown by Demiray and Kim [49]. In the remainder of this section we survey some of the approaches which could be combined in complementary experiments to study nucleate boiling as well as other phase change processes such as condensation and evaporation.

## 2.1 Contact techniques

The surface superheat temperature at the onset of nucleation on a superhydrophobic surface can be well within the measurement uncertainty of a thermocouple [1], and boiling sub-processes with drastically different physical nature could be only several micrometers apart [Moghaddam and Kiger, 2009]. Thus, techniques with high temperature measurement accuracy as well as high spatial and temporal resolutions are needed to quantify phase change processes on a boiling surface. Spatially resolved characterization of the temperature field at the solid surface can be accomplished with a number of techniques. Arrays of microfabricated electro-thermal transducers have been employed for surface temperature measurements in many phase change heat transfer experiments. For example, an array of ~100 μm wide parallel long thin film Au lines were employed to measure the evolution of the surface temperature distribution when a water drop slowly evaporated from the surface kept at 60 K before the drop impinged [1]. Because each line only measured the average temperature along the line, the results were analyzed with a numerical simulation to obtain the radial distribution of the surface temperature under the drop. In addition, microcantilevers with an integrated resistive heater and thermometer have been employed for measuring the boiling curve of microjets [2]. However, there is still ample room to push to the limits in terms of both the spatial and temperature resolutions of these microfabricated temperature sensors.

Highly sensitive measurements of heat conduction in individual nanostructures have been achieved using a number of techniques [4]. With the use of microfabricated suspended thin film Pt serpentine line resistance thermometers and sophisticated differential measurement schemes, temperature and heat flow resolutions better than $1 \times 10^{-3}$ K and $10^{-10}$ W/K have been demonstrated [5,6]. In addition, nanoscale thermocouple junctions have been fabricated for scanning thermal microscopy measurements of surface temperature distribution [7]. Continuous efforts along this direction during the past decade have allowed for quantitative surface temperature measurements with sub-100 nm spatial resolution and temperature resolution on the order of 1 K [8,9]. In a recent work that has pushed the limit of this nanoscale thermocouple based measurement technique, a temperature resolution approaching $1 \times 10^{-3}$ K was achieved with a unique modulation and averaging approach [10]. Nevertheless, these microfabricated electro-thermal transducers are contact sensors and often suffer from a long thermal time constant that is insufficient for observing the $10^{-3}$ s scale dynamics of phase change processes. Furthermore, these are intrusive measurements that can greatly affect the dynamics of these processes. This has motivated the development non-contact techniques, some of which are described below.

## 2.2 Non-contact techniques for surface temperature measurements

A variety of optical techniques have been employed for thermal measurements of phase change heat transfer. For example, infrared thermometry techniques are commonly used for measuring the solid surface temperature during phase change [11]. Silicon is transparent in the infrared spectrum, making it convenient for infrared thermal imaging of phase change heat transfer processes on surfaces fabricated on a silicon wafer. Under proper calibration procedures, this technique can be used to clearly detect the liquid-vapor-solid triple contact line under even small temperature differences (~1°C) [12]. However, the spatial resolution of such measurements is typically limited by far field diffraction to be on the order of the infrared wavelength, about 10 microns, which is often insufficient for resolving the detailed temperature distribution on nanostructured surfaces being explored to enhance boiling or condensation. In addition, the relatively low temporal resolution of this measurement approach limits it ability to capture fast transient events that require several kHz temperature sampling rate, as described later in this study.

Spatial resolution can be improved with optical techniques in the visible spectrum, such as micro-Raman spectroscopy based thermometry techniques that have been increasingly used to study heat conduction in nanostructured materials and devices [13]. Silicon, III-V compounds, graphite and other solids possess strong and distinct Raman peaks, with the line width broadened and peak position shifted with the temperature because of scattering of Raman active optical phonons with intermediate and low frequency phonons. However, the peak shift can also be caused by strain, which can result in uncertainty in the temperature determined from the peak shift. In comparison, the optical phonon temperature can be measured unambiguously from the intensity ratio of the Stokes to anti Stokes peaks. Nevertheless, accurate measurements of the anti-Stokes peak intensity may require high sample temperature. These factors have often limited the temperature resolution of micro-Raman thermometry technique to be several Kelvins or even tens of Kelvins, whereas a spatial resolution of 0.5 micron is achievable. An advantage of the micro-Raman thermometry technique is that different materials yield distinctly different Raman peak frequencies, so that the temperature difference between two ultrathin layers in contact can be measured from the respective Raman peaks [14]. This capability may allow for highly localized measurement of surface heat flux during phase change heat transfer.

Similar to Raman spectroscopy, Brillouin light scattering (BLS) is an inelastic light scattering technique. While Raman peaks arise from inelastic scattering of photons and high-frequency optical phonons, BLS peaks origin from that between photons and low-frequency acoustic phonons or magnons that are energy quanta of spin waves in ferromagnetic materials [15]. Although BLS phonon signal is usually weak, it has been recently shown that the BLS phonon peak intensity of glass depends nearly linearly with temperature [16], because the Bose-Einstein distribution of the low-frequency acoustic phonon population is reduced to the classical limit at room temperature. A temperature resolution of a few Kelvins has been demonstrated based on the BLS phonon peak intensity of glass. Moreover, the BLS magnon peak intensity can be very strong on ferromagnetic materials, with its position shifted with temperature because of temperature dependence of the exchange constant and saturation magnetization. A temperature resolution on the order of 1K has been achieved based on the BLS magnon peak position measured on a sub-100 nm thick permalloy film, whereas the spatial resolution is 0.5 micrometer [16]. Hence, BLS combined with one or two thin ferromagnetic coatings near the surface can potentially allow for sensitive non-contact measurement of the surface temperature or heat flux distribution with high spatial resolution. Similarly, time-resolved ellipsometry has been used to assess the thermal accommodation coefficient between a solid-gas interface by measurement of the photoacoustics at the Brillouin frequencies [17]. Systems consisting of inert gases (Xe, Kr,

Ar and Ne), tetrafluoroethane (R-134a), methanol and water vapor interacting with both a hydrophilic and hydrophobic Au substrate were characterized.

Sub-10 nm spatial resolution has been achieved with a thermal scanning electron microscopy (ThSEM) technique [18], where the electron diffraction pattern of Si and other crystals is used to obtain the surface temperature. Meanwhile, environmental scanning electron microscopy (ESEM) technique has been increasingly used for studying dropwise condensation [19,20]. Further development of multifunctional electron microscopy instruments may potentially allow high spatial resolution surface temperature mapping during phase change heat transfer process. Nevertheless, the temporal resolution of the ThSEM and the aforementioned inelastic light scattering techniques need to be improved by orders of magnitude in order for them to capture the fast dynamics of phase change processes.

## 2.3 Measurements of the liquid and vapor phases

The temperature distribution in the liquid and vapor phase are critical for understanding the phase change heat transfer process, yet options for temperature measurement in the liquid and vapor phase are rather limited. There have been a number of reports of using laser-induced fluorescence to measure the temperature of droplets. For example, laser induced fluorescence was used to measure the temperature of impinging droplets mixed with fluorescence dye [22]. Emission spectrum from Pyrene dye stabilized with a cationic surfactant, cetyldimethylbenzylammonium chloride (CDBAC), is used as the thermometer. The surfactant is used to prevent pyrene from grouping into micelles. Given the controversy on thermal properties of nanofluids that are liquid suspension of solid nanoparticles [23], it remains to be better understood whether the fluorescence dye can affect the phase change processes of the liquid and vapor phases. Similarly, thermochromic liquid crystal (TLC) thermography has been used to measure the temperature near the triple line in the evaporation front of a water filled microcapillary [24]. However, this technique has a very limited temperature range (~4°C) and the same concerns raised by the presence of long polymeric molecules in fluorescence techniques apply here with the introduction of the liquid crystals.

At MIT, researchers have expanded the infrared (IR) thermography approach by synchronizing it with high-speed video (HSV) from below and beside the boiling surface, and Particle Image Velocimetry (PIV) [50,51]. Also, the DEtection of Phase by Infrared Thermometry (DEPIcT) technique was developed, which uses the differences in emissivity of the liquid and vapor phases to measure the phase distribution (including the shape of the microlayer) on the boiling surface of IR-transparent heaters [52,53]. Examples of the data obtained at MIT from these two approaches are shown in Figures 1-4, and details are given in the figure captions.

It is also desirable to develop other non-contact optical temperature measurement techniques without the need of seeding temperature sensitive materials in the heat transfer fluids. For example, a Global Rainbow Thermometry has been developed to measure the temperature of water droplets based on the temperature dependence of the refractive index and light scattering pattern of the water droplet [25]. Another possibility is the use of near infrared (NIR) absorption thermometry which has been used to measure water temperature in PDMS microchannels with a resolution of 0.2K [26]. These and other non-invasive techniques represent a research direction worth further investigations.

## 3 Measurement at the single bubble level.

Having discussed a variety of techniques for measuring the solid, fluid, and vapor phases, we now turn to an experiment designed to study the nucleation and departure of single bubbles. Measurement at the single bubble level is a key to deciphering the physics of the complex heat and mass transfer processes involved in boiling, and could pave the way for better understanding of the higher order effects resulting from multiple bubbles interactions. While it is common knowledge that bubble generation at the surface is responsible for the observed enhanced heat transfer in boiling, details of the heat transfer processes triggered by bubble formation and departure are not clearly understood. Different and often contradictory hypotheses have been proposed to describe the nature of the heat transfer processes [54-65], and although great efforts have been devoted to development and testing of these models, the necessary measurement tools have not been generally available to test the basic assumptions of the models at the single bubble level. This is because performance of the models has often been evaluated based on their ability to predict the overall surface heat transfer coefficient, which is the cumulative effect of all microscale boiling subprocesses. While this bulk measure is often the quantity of engineering interest, such a simple validation metric is insufficient to answer why a particular model fails.

Figure 5 shows the top view of a device that generates single bubbles and determines the surface temperature and heat flux at the surface-bubble interface using three layers of temperature sensors imbedded within a composite wall. The details of the device are discussed in Refs. [66-69]. The device die is attached to a Pin Grid Array (PGA) that is installed on the bottom of a test liquid chamber and is connected from below to a custom-made Signal Conditioning Board (SCB) through a stack of sockets. The output of the SCB is directly connected to an A/D board installed in a PC. The temperature sensors are calibrated with an accuracy of $\pm 0.1$ °C. Data are collected during boiling of FC-72 liquid at saturation conditions under atmospheric pressure. Figure 6 shows images of a bubble at a surface temperature of 80.2 ºC, and Fig. 7 shows the surface temperature and heat flux results corresponding to the bubbling event.

Comparison of the bubble images and the temperature data show that the initial formation of the bubble at t = 3.8 ms was associated with a sudden drop in surface temperature. The temperature drop started at the center of the array and progressed over the subsequent sensors. A second phase of surface temperature decline started after the bubble/surface contact area reached its maximum diameter and the apparent contact line started to advance over the contact area. The advancing liquid rewetted the dried out area contact area. This rewetting process began at about t = 5.8 ms. The process resulted in a continued decrease in temperature of sensor S-5 that had already significantly decreased due to the prior cooling event. The temperature decrease trend passed as a radially inward moving wave, corresponding to when the contact line successively passed over sensors S-4 to S-1. As can be seen in the lower panel of Fig. 9, the surface temperature outside the contact area remained unchanged during the entire bubble growth and departure process. Comparison of the bubble contact radius with the surface temperature history showed that the temperature drop at each sensor started after the apparent contact line passed over the sensor. This suggested that the observed temperature drop was due to surface cooling resulted from evaporation of a thin liquid layer (i.e., a "microlayer") left over the surface, after the contact line rapidly receded. The surface temperature started to increase shortly after the microlayer was mostly evaporated. These data clearly reinforce the results of prior studies [70-74] regarding the microlayer presence underneath a bubble, and suggest that the process is active primarily at the bubble/surface contact area, in contrast with the Mikic and Rohsenow [58] model that suggested an active area of twice the bubble diameter, and much closer to the more recent observations of Yaddanapudi and Kim [75]. Mikic and Rohsenow [58] cited Han and Griffith [34] as a basis for their assumption. However, Han and Griffith [76] did not provide a solid reason or any particular experimental evidence for their assumption. They simply assumed that following the departure of a bubble from the heating surface, a volume element of superheated liquid from an area twice the bubble diameter is brought into the bulk fluid.

The example results presented here suggest that in order to fully capture the heat transfer events at the bubble-surface interface, a temperature sampling rate on the order of 10 kHz is required. So, thermal imaging with 1-2 orders of magnitude lower sampling rate does not seem to be a suitable tool for capturing the details of the boiling sub-processes, although such tools can be used to capture the integral effect of the cooling events over a period of time. In addition, temperature sensors and the heated wall structure/properties should be carefully designed to isolate and study fast events (such as the microlayer evaporation process). For example, thermally thick walls (this is relative to the boiling characteristics of a particular fluid) do not allow proper isolation of different heat transfer events.

## 4. Molecular- and nanoscale resolution of interfacial phenomena

There are few experimental methods that are capable of resolving the presence of a few molecular layers of liquid condensed on a surface. These adsorbed layers are critical, however, for determining the surface energies and the kinetics of evaporation and condensation. Sum-frequency vibrational spectroscopy can be used to probe the vibrational states at the interfaces but requires complex and expensive equipment and nearly perfect surfaces to suppress diffuse light scattering. X-ray or neutron reflectivity are also capable of directly resolving thin adsorbed layers but require large facilities to carry out the work and the sample requirements are highly restrictive. Photothermal and photoacoustic phenomena provide more generally applicable methods for probing the thermodynamics and kinetics of surfaces under conditions that are relevant for heat transfer applications. In the pump-probe ellipsometry approach, a short pump laser pulse creates a rapid temperature excursion of a surface and a time-delayed probe laser pulses measures the propagation of an acoustic wave in the vapor that is generated by heat transfer from the surface to the vapor [27] as well as the exchange of mass between the surface and vapor [17]. The time-resolved ellipsometry experiments revealed more extensive adsorption of R134a refrigerant on Au surfaces that were coated with OH-terminated self-assembled monolayers than on Au surfaces with CH4 termination [17]. The data show that while the macroscopic contact angle of refrigerants is nearly constant at very small values, the free energies of nanoscale adsorbed layers can nevertheless be affected by modifying surface chemistry.

The free energy of liquid-vapor interfaces creates a large nucleation barrier for the formation of a vapor bubble within a homogeneous liquid. At a solid surface, the nucleation barrier for the formation of vapor is reduced but homogeneous nucleation rates remain low. Therefore, rapid phase change of liquid-to-vapor in a heat-transfer application is usually thought to require pre-existing vapor-liquid interfaces that circumvent the slow nucleation kinetics. At Illinois, researchers have investigated the evaporation rate of water droplets that adhere to hydrophilic surfaces [28] and the time-scale for the formation of a vapor layer during the droplet rebound from a hydrophobic surface [29]. In both cases, a small drop of liquid was brought rapidly into contact with a hot surface, and high frequency thermoreflectance measurements were used to determine when vapor forms at the solid-liquid interface with an ultimate time resolution on the order of 10 microseconds. A novel optical thermometry—based on the temperature dependence of two-photon absorption in an indirect gap semiconductor—enabled temperature measurements with similarly high time resolution. Measurements of temperature of Si using this approach have a spatial resolution on the order of 10 microns and a noise floor on the order of 1 K in a 1 kHz bandwidth [30].

The apparatus used for the thermoreflectance measurements is shown in Figure 8, while Figure 9 shows illustrative transient temperature and effective thermal interface conductance data obtained during the impingement of dispensed water droplets at different surface temperatures. Because of the transient process of the impinging drops, the measurements of the transient absorption signal were synchronized to the trigger of drop dispenser. The time offset between the measurement and trigger of the dispenser was varied between 0 and 1500 ms. Results from a number of measurements made at the same time offset were averaged to obtain the evolution of the surface temperature and thermal conductance as a function of time after drop impingement on the substrate. The time evolution data were analyzed to obtain the heat transfer and residence time of the drop evaporation process, and it was found that the heat flux could exceed 500 W/cm$^2$ for a short duration on the order of 10 ms when the surface temperature was increased near 200 $^o$C.

## 5. Conclusions

We have seen that there are many specialized measurement methods for studying phase change heat transfer, and that there is still much that can be learned from data generated today with state-of-the-art diagnostics. Indeed, transformative improvements in the understanding, prediction and control of phase change heat transfer may come from the development of new mechanistic models and direct numerical simulations, with inspiration and validation coming from a combination of existing experimental techniques that can probe temperature and heat flux with sufficient sensitivity over multiple length and time scales. However, there is great room for progress in experimental methods, and continuation of the following trends is advocated: i) more systematic use of combined (synchronized) diagnostics, to develop a complete physical picture of phase change phenomena; ii) development of temperature and velocity sensors with ever higher temporal resolution (>5 kHz) and spatial resolution (<5 μm), particularly useful as we start to understand how phenomena at the microscale (e.g., microlayer evaporation) can affect macroscopic figures of merit such as the heat transfer coefficient and the Critical Heat Flux (CHF); and iii) more systematic use of engineered surfaces (engineered at the micro- and nano-scale), to control the composition and texture of the boiling surface, thus enabling investigation of the effects of roughness, wettability, porosity, presence of cavities, size and shape of cavities, and surface thermo-physical properties on boiling heat transfer.

We are confident that advances in experimental methods will significantly impact the discipline of phase change heat transfer, including the rapidly expanding research in phase change heat transfer on nanostructured surfaces. We emphasize, however, that further development of the field also requires clear accounting of what is known and what is unknown, i.e., that scientific progress requires the formulation of thoughtful questions that the field will ask and then subsequently answer as new capabilities for experiment come on-line. Theory and computational models typically require input parameters that are determined from experiment, and theory often includes simplifying assumptions that must be validated by experiment. This interplay between theory, modeling, and experiment will be critical for a better fundamental understanding of the key physical processes in phase change heat transfer, and for the translation of this understanding into improved performance in engineering applications.

## FIGURE CAPTIONS

**Figure 1:** (a) Schematic diagram of the experimental configuration, not to scale. The Al heater block is 1 cm thick and the diameter of the hole in the Al heater block is 1.5 cm. Droplet generator, sample, and objective lens are vertically aligned. (b) The sample is a 1 mm thick double-side polished Si wafer. The bottom film of $TiO_2$ is an antireflection coating and the top film of $TiO_2$ thermally isolated the Ti thin film from the Si wafer. The Pt film provides a chemically inert surface, stable against boiling water. (c) Optical layout of the pump-probe system used. Sample region is described in detail in Fig 1(a).

**Figure 2:** Illustrative data showing transient changes of temperature and effective thermal conductance created by the impingement of dispensed water volume of 0.19 $mm^3$. Panels (a) and (c) are for a relatively low sample temperature of 130 °C. Panels (b) and (d) are for a relatively high sample temperature of 210 °C. Time zero is defined by the electronic trigger of the microdispenser. The series of water droplets arrives at the sample surface 50 ms after the trigger.

**Figure 3:** Representative synchronized PIV and high-speed video images of a growing and departing steam bubble at atmospheric pressure. (from [51])

**Figure 4:** Post-processing of the PIV data in Figure 3 yields the values of the z-component of the vorticity vector in the wake of a steam bubble. The abscissa represents time after bubble nucleation (bubble departure occurs at t=16 ms). Maximum uncertainty for vorticity values is ~18%. (from [51])

**Figure 5:** Phase distribution on the boiling surface at various values of heat flux (normalized to CHF) and subcoolings, obtained with eth DEPIcT technique. Black represents dry regions, grey and white represent wet regions. The fluid is water at atmospheric pressure, boiling over horizontal silicon heaters. Note that the subscript 'CHF(-)' indicates the last heat flux step before CHF occurs.

**Figure 6:** Post-processing of the DEPIcT data in Figure 5 yields the time-averaged wetted fraction (i.e. fraction of boiling surface that is contact with the liquid phase) as a function of normalized heat flux and subcooling.

**Figure 7:** Close view of the MEM device showing the topmost sensor layer radially distributed. The bottom-right inset figure shows two sensors (i.e. H-1 and H-2) made at different levels beneath the sensor array. All temperature sensors are Resistance Temperature Detector (RTD). Material of the sensors is Ni and their thickness is approximately 10 nm. The H-1 and H-2 sensors are coil-shape with a diameter of 1 mm. Spacing between the H-1 and H-2 sensors is 7.5 µm. The sensor array is 2.5 µm above the H-2 sensor. The sensor array is covered with a 0.2-µm thick polymer layer. The top-right SEM image shows 0.7, 1.3, and 2.4 µm in diameter cavities.

**Figure 8:** A bubbling event at surface temperature 80.2 °C. A waiting time of 2.9 ms exists between the bubbles. Time is in millisecond.

**Figure 9:** Surface temperature (upper panel) and heat flux (lower panel) variations during the bubbling event. Heat transfer during microlayer evaporation and transient conduction processes are marked. S-1 through S-8 indicate sensors 1-8 in Figure 7.

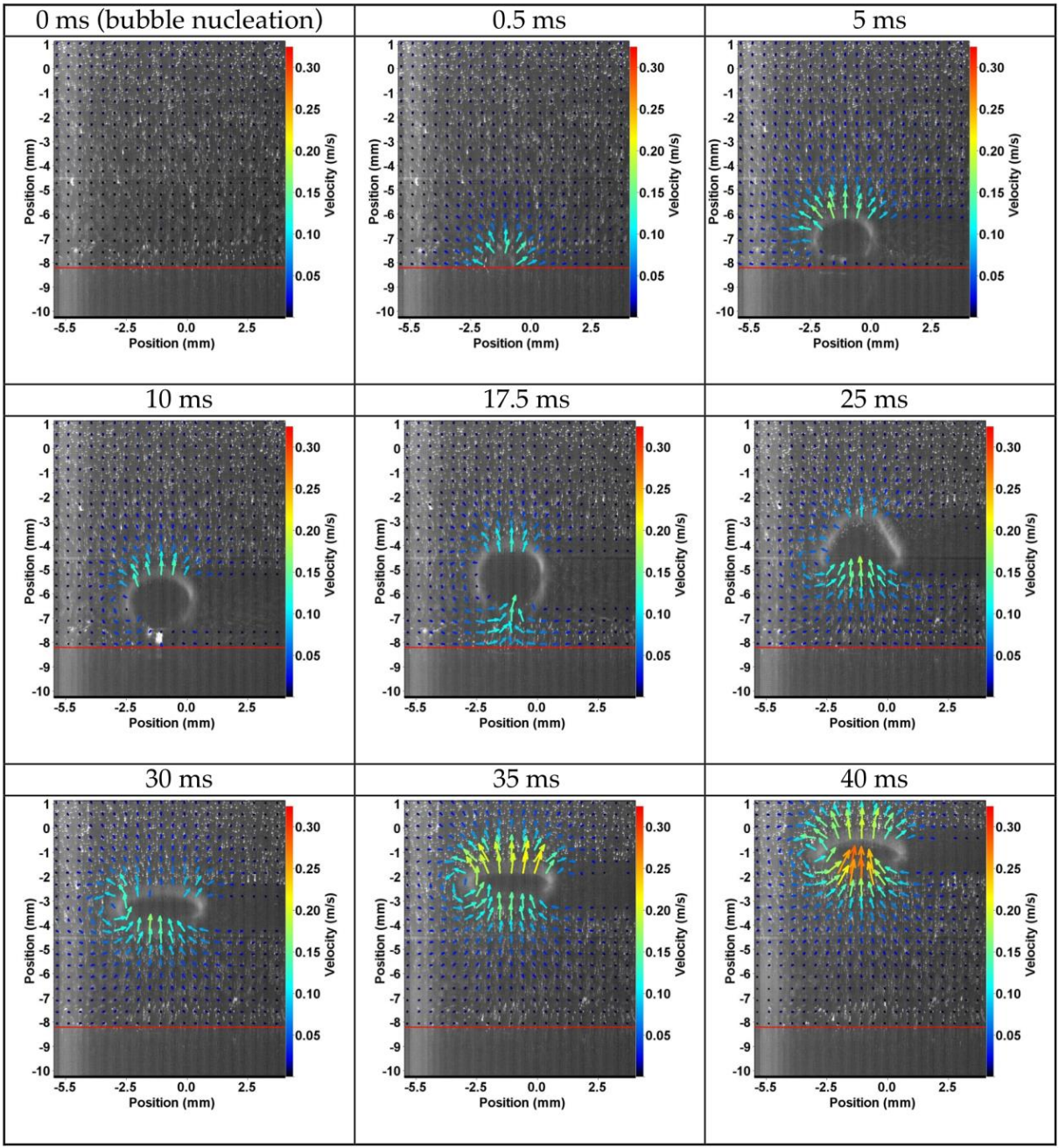

**Figure 1:** Representative synchronized PIV and high-speed video images of a growing and departing steam bubble at atmospheric pressure. (from [51])

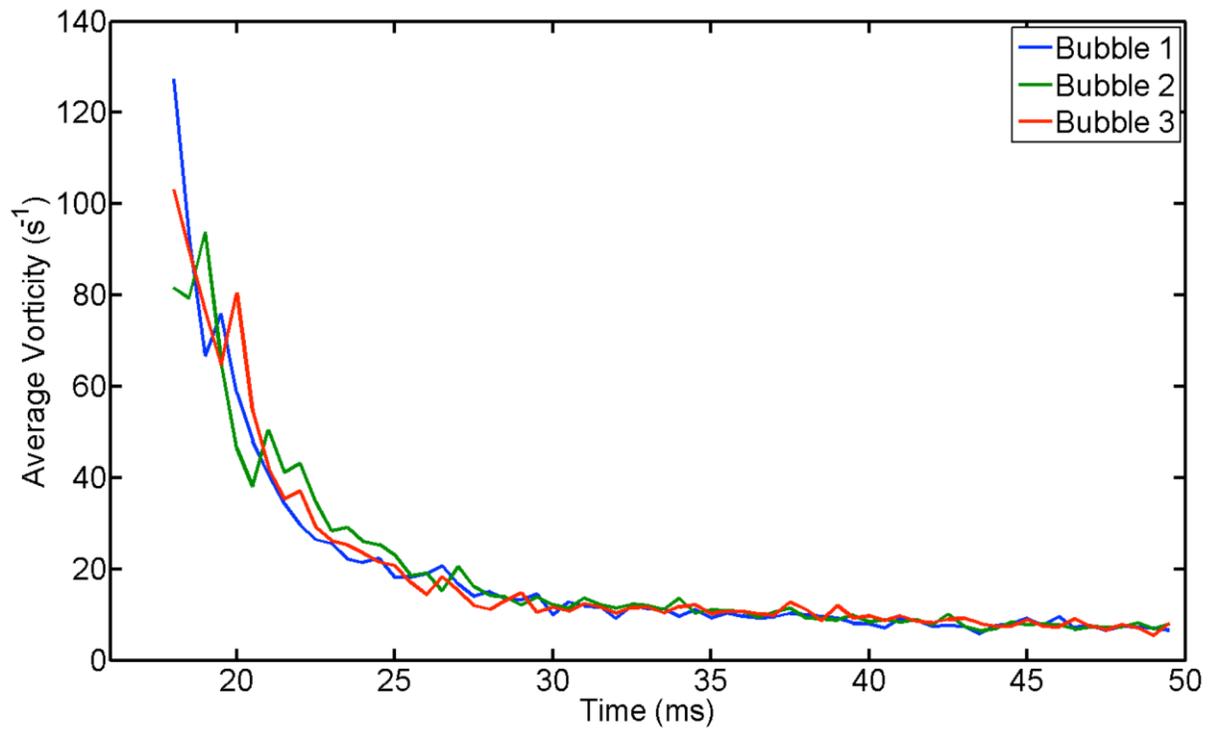

**Figure 2:** Post-processing of the PIV data in Figure 3 yields the values of the z-component of the vorticity vector in the wake of a steam bubble. The abscissa represents time after bubble nucleation (bubble departure occurs at t=16 ms). Maximum uncertainty for vorticity values is ~18%. (from [51])

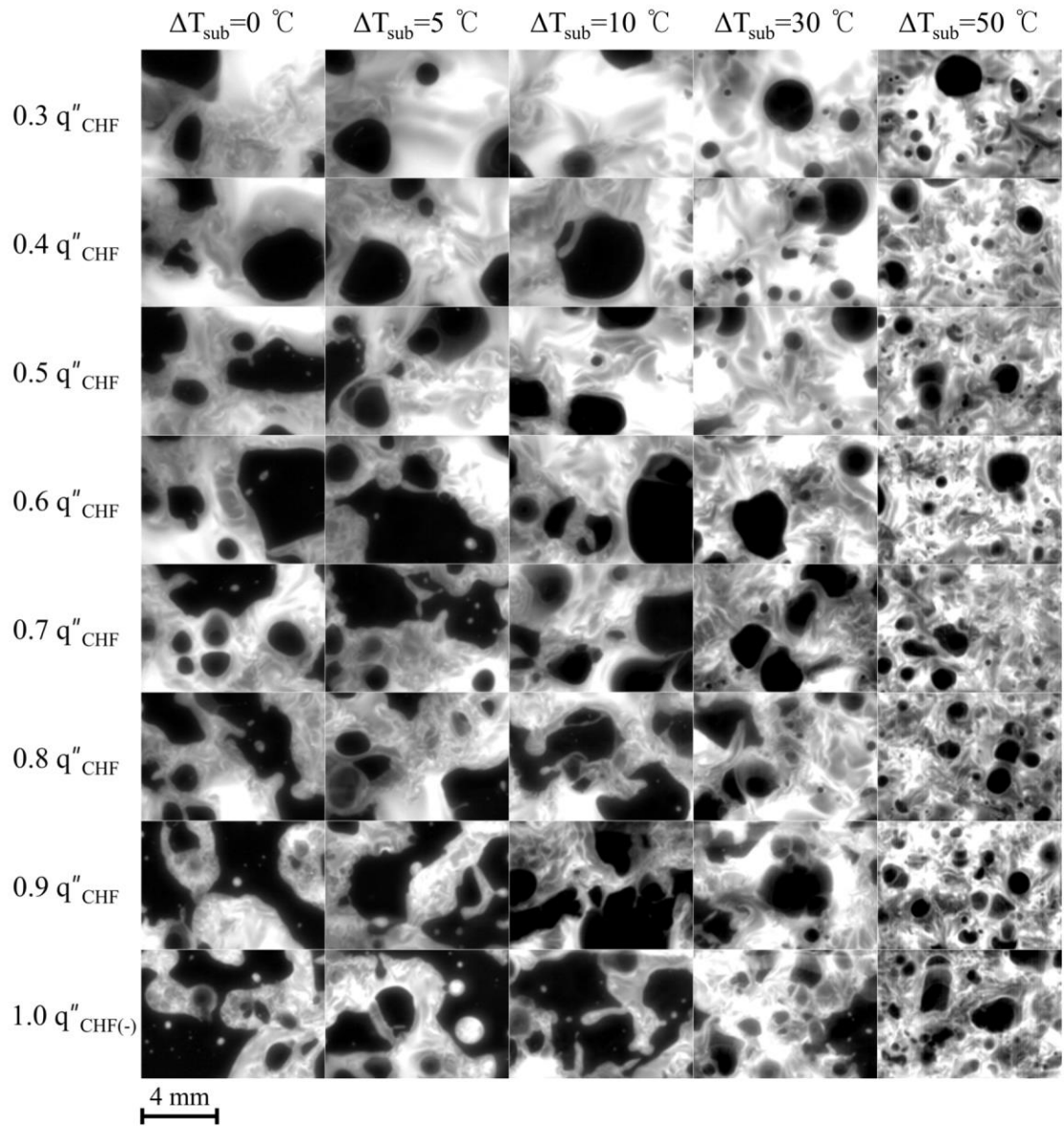

**Figure 3:** Phase distribution on the boiling surface at various values of heat flux (normalized to CHF) and subcoolings, obtained with eth DEPIcT technique. Black represents dry regions, grey and white represent wet regions. The fluid is water at atmospheric pressure, boiling over horizontal silicon heaters. Note that the subscript 'CHF(-)' indicates the last heat flux step before CHF occurs.

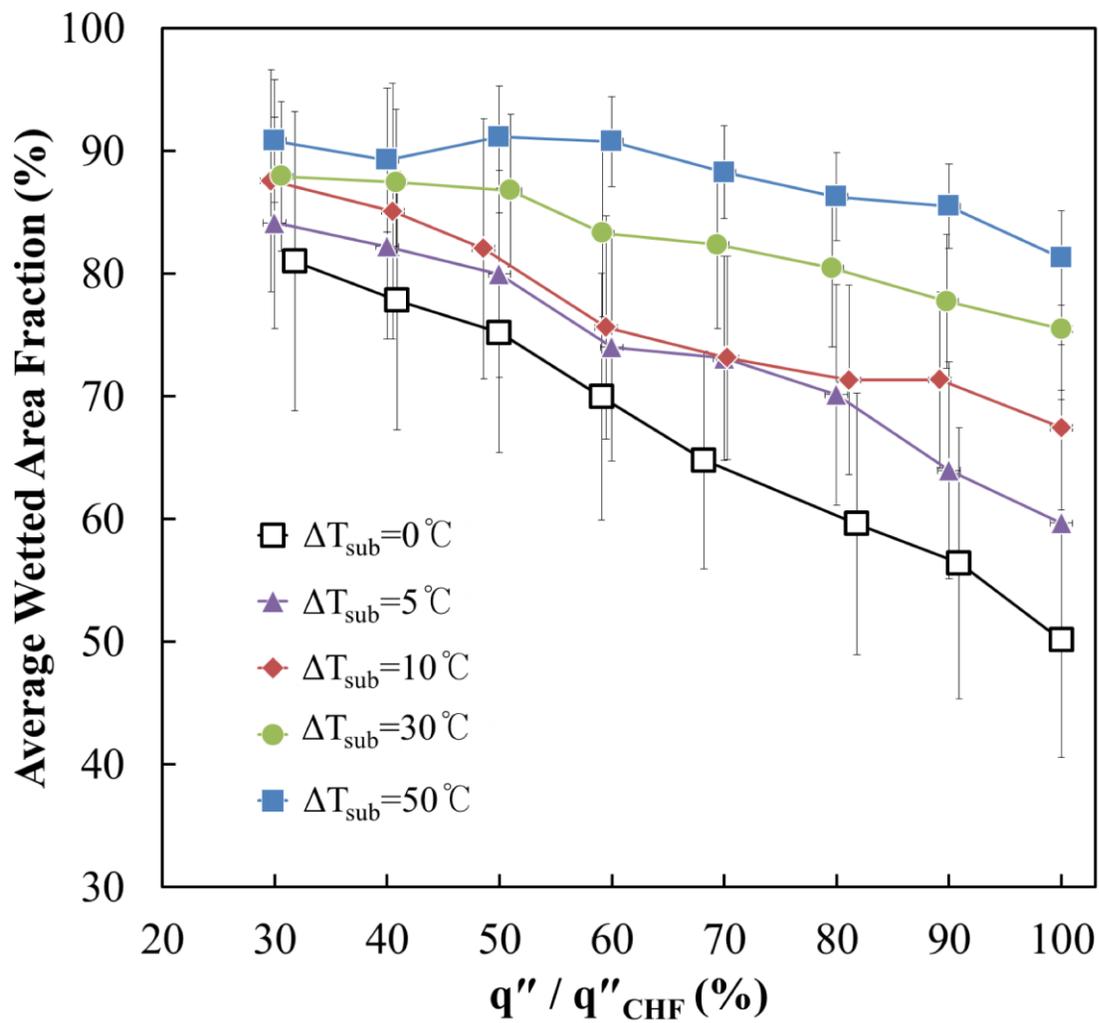

**Figure 4:** Post-processing of the DEPIcT data in Figure 5 yields the time-averaged wetted fraction (i.e. fraction of boiling surface that is contact with the liquid phase) as a function of normalized heat flux and subcooling.

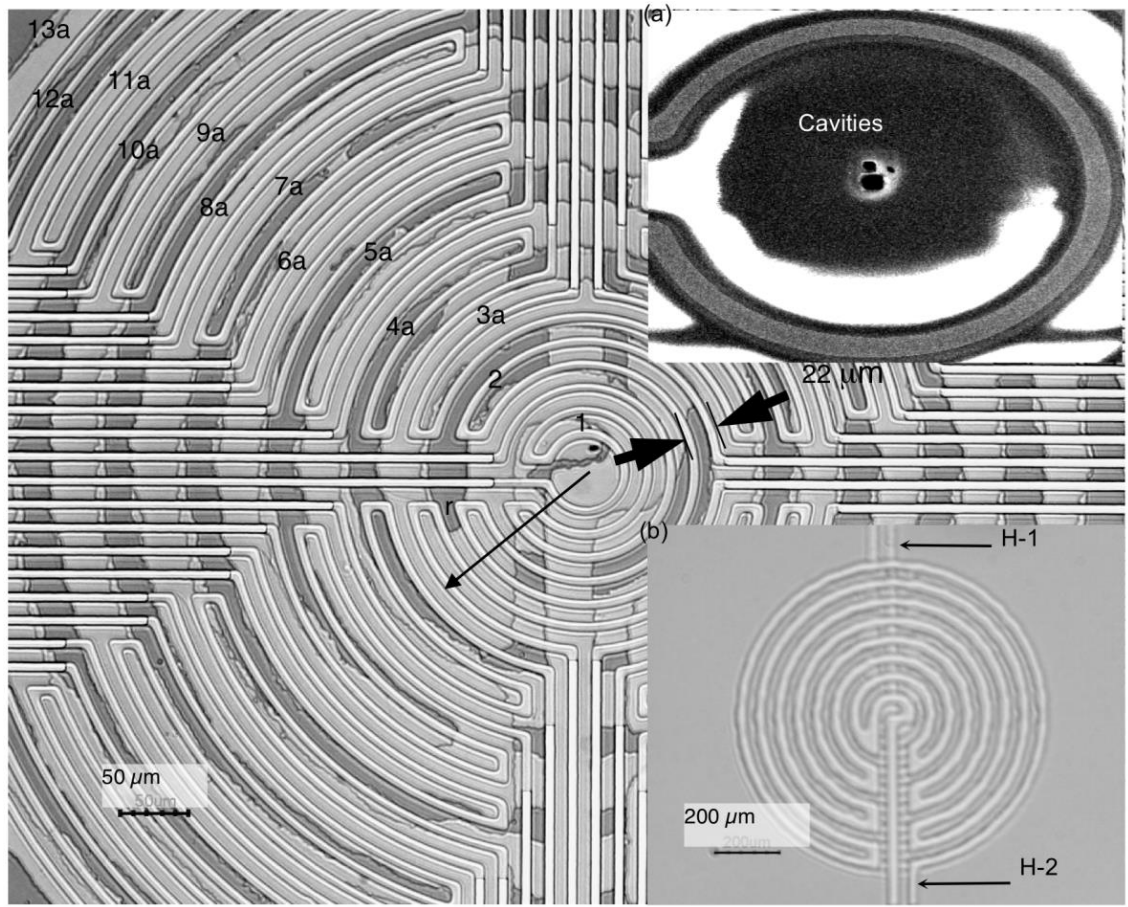

**Figure 5:** Close view of the MEM device showing the topmost sensor layer radially distributed. The bottom-right inset figure shows two sensors (i.e. H-1 and H-2) made at different levels beneath the sensor array. All temperature sensors are Resistance Temperature Detector (RTD). Material of the sensors is Ni and their thickness is approximately 10 nm. The H-1 and H-2 sensors are coil-shape with a diameter of 1 mm. Spacing between the H-1 and H-2 sensors is 7.5 µm. The sensor array is 2.5 µm above the H-2 sensor. The sensor array is covered with a 0.2-µm thick polymer layer. The top-right SEM image shows 0.7, 1.3, and 2.4 µm in diameter cavities.

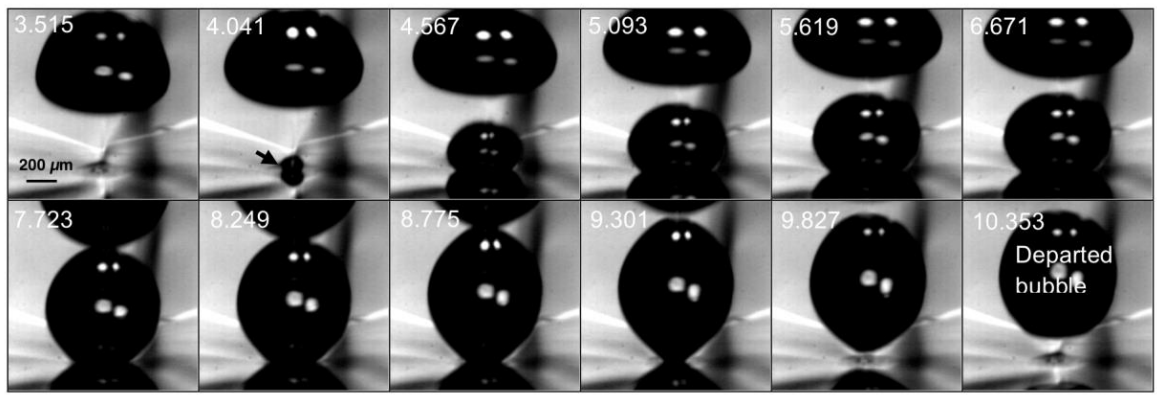

**Figure 6:** A bubbling event at surface temperature 80.2 °C. A waiting time of 2.9 ms exists between the bubbles. Time is in millisecond.

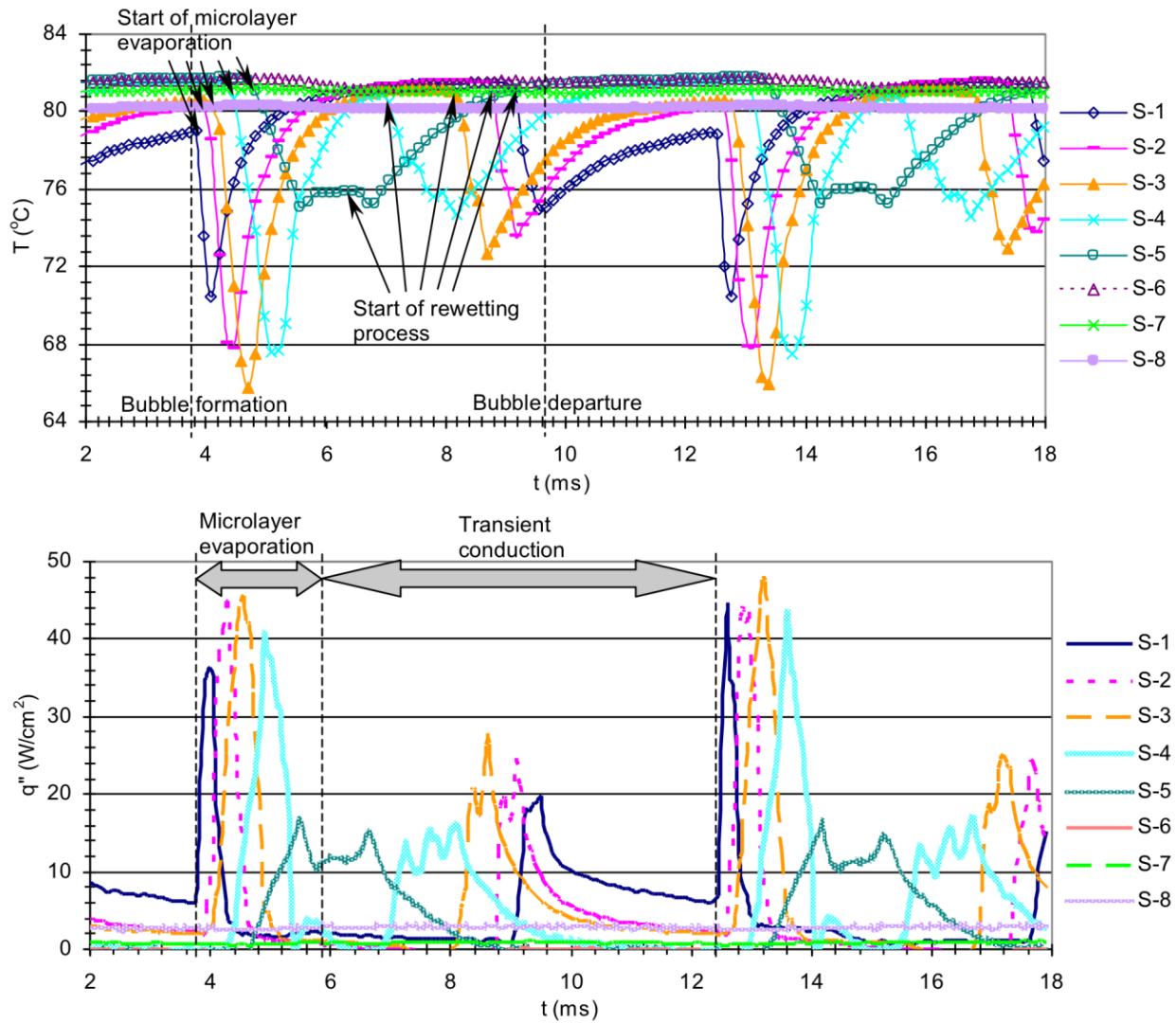

**Figure 7:** Surface temperature (upper panel) and heat flux (lower panel) variations during the bubbling event. Heat transfer during microlayer evaporation and transient conduction processes are marked. S-1 through S-8 indicate sensors 1-8 in Figure 5.

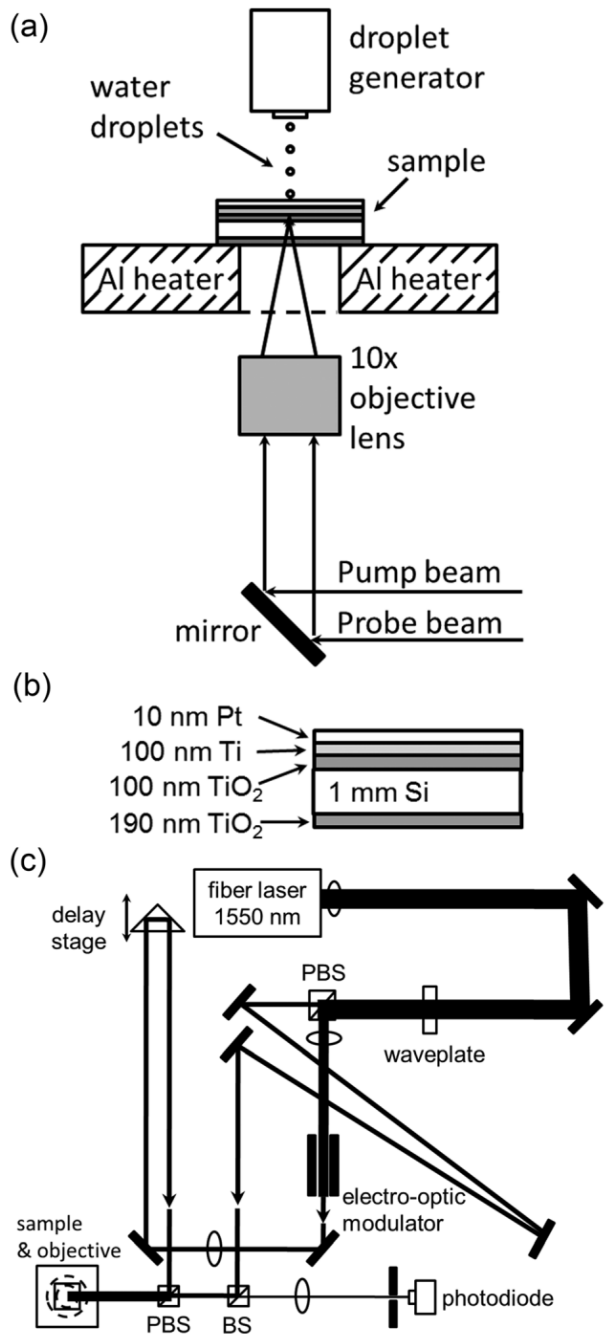

**Figure 8:** (a) Schematic diagram of the experimental configuration, not to scale. The Al heater block is 1 cm thick and the diameter of the hole in the Al heater block is 1.5 cm. Droplet generator, sample, and objective lens are vertically aligned. (b) The sample is a 1 mm thick double-side polished Si wafer. The bottom film of $TiO_2$ is an antireflection coating and the top film of $TiO_2$ thermally isolated the Ti thin film from the Si wafer. The Pt film provides a chemically inert surface, stable against boiling water. (c) Optical layout of the pump-probe system used. Sample region is described in detail in Fig 1(a).

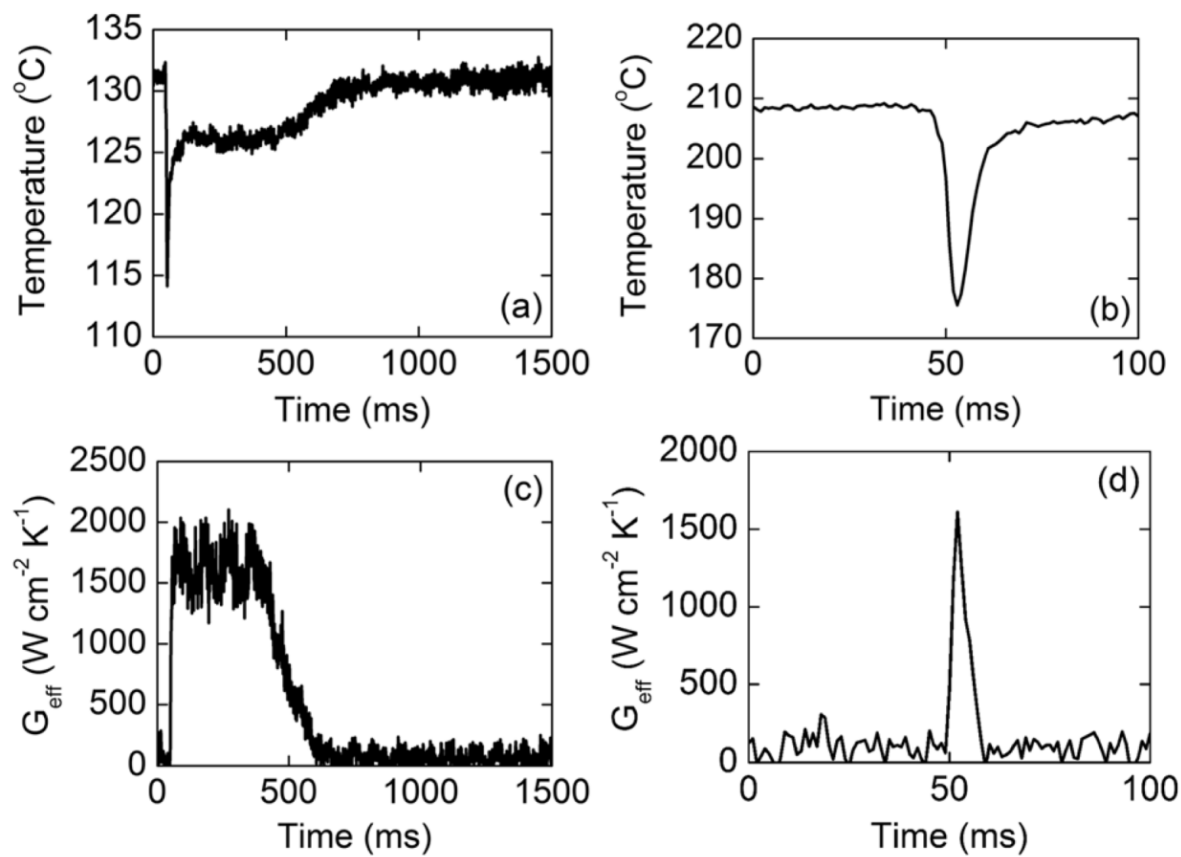

**Figure 9:** Illustrative data showing transient changes of temperature and effective thermal conductance created by the impingement of dispensed water volume of 0.19 mm³. Panels (a) and (c) are for a relatively low sample temperature of 130 °C. Panels (b) and (d) are for a relatively high sample temperature of 210 °C. Time zero is defined by the electronic trigger of the microdispenser. The series of water droplets arrives at the sample surface 50 ms after the trigger.